\documentclass[reprint,amsmath,amssymb,aps,pra,floatfix,showkeys, nofootinbib]{revtex4-2}
\usepackage{graphicx}
\usepackage{bm,mathtools,booktabs,microtype}
\usepackage[colorlinks=true,linkcolor=blue,citecolor=blue,urlcolor=blue]{hyperref}
\usepackage{comment}
\newcommand{\sfR}{\mathsf{R}}
\newcommand{\sfPi}{\mathsf{\Pi}}

\newcommand{\calD}{\mathcal{D}}
\newcommand{\dd}{\mathrm{d}}
\newcommand{\tr}{\operatorname{tr}}
\newcommand{\diag}{\operatorname{diag}}
\newcommand{\half}{\frac{1}{2}}

\newcommand{\RR}{\mathbb{R}}

\begin{document}
\title{Elementary-Symmetric Ricci Gravity: Metric Formulation}
\author{Xingpao Suo}
\email{xpsuo@zju.edu.cn}
\author{Dianxu Ma}
\affiliation{College of Physics and Information Engineering, Zhaotong University, Zhaotong, Yunnan 657000, China}
\author{Yuqiong Shan }
\affiliation{College of Physics and Information Engineering, Zhaotong University, Zhaotong, Yunnan 657000, China}
\author{Xingyan Suo}
\affiliation{Turpan Gaochang District Secondary Vocational and Technical School, Turpan, Xinjiang, China}
\date{\today }
\begin{abstract}
We investigate a four-dimensional metric theory of gravity whose Lagrangian is a linear combination of the elementary symmetric polynomials of the Ricci tensor. We derive the field equations and characterize Einstein vacua and a real, diagonalizable class of parallel-Ricci vacua. The latter includes direct products of two constant-curvature surfaces and can exhibit a hierarchy of curvature radii. The Einstein vacua possess multiple branches associated with the distinct roots of a polynomial, whose multiple roots define boundaries in parameter space. We derive the linearized equations on maximally symmetric backgrounds, distinguish the scalar and spin-2 degrees of freedom, and then analyze their classical linear stability. We also obtain the generalized Friedmann constraint and the associated homogeneous evolution equations. An illustrative spatially flat solution evolves from the neighborhood of an unstable de Sitter solution toward a stable de Sitter solution. This behavior may motivate further investigation into early- and late-time accelerated expansion and provide a gravity-driven mechanism for automatic  end of inflation. 
\end{abstract}
\keywords{higher-curvature gravity, Ricci invariants, de Sitter vacua, cosmological dynamics}
\maketitle

\section{Introduction}
Higher-curvature gravity (HCG) theories are natural extensions of general relativity (GR).
There are at least three motivations for introducing HCG theories. First, they provide a framework for investigating the ultraviolet behavior of gravity~\cite{Stelle1977}. For instance, quadratic curvature terms improve the perturbative ultraviolet behavior, although their additional propagating modes raise important questions concerning unitarity and stability~\cite{Stelle1977,Stelle1978}.
Second, HCG theories can provide alternatives to GR-based cosmological models~\cite{NojiriOdintsov2011}. Compared with GR, they may offer advantages in certain respects, particularly by providing dynamical mechanisms for accelerated expansion in the early and late Universe and, in specific models, by avoiding the Big Bang singularity~\cite{Starobinsky1980,NojiriOdintsov2011}. 
Finally, HCG has a natural interpretation within gravitational effective field theory, although, like GR, HCG theories need not be regarded as fundamental theories at the quantum level~\cite{Donoghue1994}.

Different classes of HCG theories can be constructed from distinct motivations or formalisms~\cite{NojiriOdintsov2011}. For example, $f(R)$ theories can be constructed to avoid certain ghost modes~\cite{SotiriouFaraoni2010,DeFeliceTsujikawa2010}; Lovelock gravity~\cite{Lovelock1971} is constructed primarily to formulate higher-curvature Lagrangians in higher dimensions while keeping the resulting field equations second order in metric derivatives and avoiding corresponding higher-derivative instabilities; and nonlocal gravity~\cite{Biswas2012} is mainly constructed to resolve spacetime singularities while avoiding the ghost instabilities that can arise in finite higher-derivative gravity.

In this paper, we study a metric theory organized by the elementary symmetric polynomials of the Ricci endomorphism. We refer to this construction as elementary-symmetric Ricci gravity (ESRG).
The distinctive feature of ESRG lies in its systematic organization of curvature invariants through the characteristic polynomial of the Ricci endomorphism. Rather than introducing arbitrary combinations of higher-order curvature scalars, the theory employs the elementary symmetric polynomials as a natural algebraic basis. This construction provides a unified framework for studying the nonlinear curvature dependence of gravitational dynamics, particularly the multiplicity of algebraic vacuum branches and the corresponding cosmological evolution. Related theories organized in terms of Ricci curvature polynomials have also been investigated \cite{HaoZhao2017,LiLiuLu2018}. Our focus is on the algebraic structure of the vacua, the perturbative stability, and homogeneous cosmology. 

There are trade-offs among HCG theories. For example, viable $f(R)$ gravity models can avoid a spin-2 ghost, but they introduce an additional scalar degree of freedom and can mediate a fifth force, potentially violating the strong equivalence principle; like GR, they remain nonrenormalizable~\cite{SotiriouFaraoni2010,DeFeliceTsujikawa2010}. 
Nonlocal gravity offers a promising approach to improving the ultraviolet behavior of gravitational interactions and avoiding ghostlike excitations. However, its formulation and physical interpretation remain challenging, particularly regarding causality, the initial-value problem, nonlinear dynamics, and the preservation of unitarity under quantum corrections\cite{HindawiOvrutWaldram1996a,Woodard2015}.
Like quadratic gravity, our theory also accommodates ghost degrees of freedom. Nevertheless, given that even standard GR cannot be satisfactorily quantized, we set aside quantum difficulties and confine our discussion up to the classical linear level.

From the perspective of fundamental dynamical variables, gravitational theories can be divided into two classes: metric formulations and metric-affine formulations (the latter include the Palatini formulation, which we contrast below)~\cite{HehlMcCreaMielkeNeeman1995,Olmo2011}. In metric formulations, the connection is fixed to be the Levi-Civita connection and is not a fundamental dynamical variable. By contrast, the Palatini formulation treats the metric and the connection as independent dynamical variables. The field equations are obtained by varying the action with respect to the metric and the connection separately. Therefore, for formally identical actions, one may construct metric theories and Palatini theories, respectively. For GR without matter coupling, the two formalisms are equivalent. However, this equivalence generally fails for nonlinear curvature actions~\cite{Olmo2011,BeltranJimenez2019}.
The Palatini formulation admits a richer geometric structure and avoids propagating scalar degrees of freedom in some models~\cite{Olmo2011}; yet most Palatini theories lead to more complicated matter--gravity couplings, and some models can generate curvature singularities at the surfaces of high-density polytropic matter configurations~\cite{Olmo2011,BarausseSotiriouMiller2008}. 
By contrast, while many higher-curvature metric theories introduce ghost degrees of freedom, they offer a clear geometric interpretation and constitute a standard framework for cosmological research~\cite{Stelle1977,Stelle1978}. 
For these reasons, this work focuses on the metric formulation rather than the Palatini formulation. Throughout this work, the connection is the Levi-Civita connection of the metric unless stated otherwise.

This paper is organized as follows. In Sec.~\ref{sec:model}, we introduce the model and the principles behind it. We discuss the general model and then introduce the linear model on which we focus. In Sec.~\ref{sec:FE}, we derive the field equations and analyze several algebraic vacua, including multibranch Einstein and Ricci-flat vacua. In Sec.~\ref{sec:perturbation}, we analyze linear perturbations around an (A)dS vacuum and the stability of the background. Like most HCG theories, our model has ghost degrees of freedom. 
With these results as a foundation, in Sec.~\ref{sec:FLRW}, we investigate Friedmann--Lema\^itre--Robertson--Walker (FLRW) cosmology in this model. We find that, for appropriate parameters and initial conditions, the Universe evolves from an unstable de Sitter vacuum toward an attracting de Sitter vacuum. This potentially offers a natural pure-gravity model for the automatic  end of inflation.
In Sec.~\ref{sec:discussion}, we discuss potential research directions and future research plans.

Note that throughout this paper, we use the metric signature $(-,+,+,+)$ and a curvature convention for which de Sitter space has $R_{\mu\nu}=3H^2g_{\mu\nu}$. We use natural units, in which $c=\hbar=1$.

\section{Elementary-symmetric Ricci gravity} \label{sec:model}
 Let $\sfR$ denote the matrix with elements $R^\mu{}_\nu$\footnote{In mathematics, this matrix is commonly referred to as the Ricci endomorphism, which defines a linear transformation on the tangent space at each point of the manifold.}.  The general ESRG action in $D$ spacetime dimensions is
\begin{equation}
 S[g]=\frac{1}{2\kappa}\int\dd^Dx\,\sqrt{-g}\,F(c_1,\ldots,c_D)+S_m, 
 \label{equ:equ_action}
\end{equation}
where $\kappa:=8\pi G$, $F:\RR^D\to\RR$ is a function, $S_m$ is the matter action, and the elementary symmetric polynomials $c_k=c_k(\sfR)$ are defined by the expansion 
\begin{equation}
 \det(zI-\sfR)=\sum_{k=0}^{D}(-1)^k c_k z^{D-k}.
 \label{equ:def_ck}
\end{equation}
In four dimensions,
\begin{align}
 c_0&=1, \nonumber\\ 
 c_1&=[\sfR]=R,\nonumber\\
 c_2&=\tfrac12\bigl([\sfR]^2-[\sfR^2]\bigr),\nonumber\\
 c_3&=\tfrac16\bigl([\sfR]^3-3[\sfR][\sfR^2]+2[\sfR^3]\bigr),\nonumber\\
 c_4&=\det\sfR,
 \label{equ:ck4}
\end{align}
where $[A]=\tr A$. The choice $F=c_1-2\Lambda$ reproduces general relativity with cosmological constant $\Lambda$, and $F=f(c_1)$ gives  $f(R)$ gravity.

The elementary symmetric polynomials can also be defined using
\begin{equation}
 \det(I+\ell^2\sfR)=\sum_{k=0}^{D}\ell^{2k}c_k(\sfR ).
\end{equation}
Thus, a Born--Infeld-type Lagrangian can be obtained if $F(c_1,\ldots,c_D):= \sqrt{|\sum_{k=0}^{D}\ell^{2k}c_k|}$, with the normalization and constant subtraction specified separately \cite{DeserGibbons1998}. The Lagrangian density $F(c_1,\ldots,c_D):=\sqrt{|c_D|}$ recovers an Eddington-type determinant expression, even though these algebraic identities do not identify the resulting metric theory with a purely affine Eddington theory or an Eddington-inspired Born--Infeld theory with an independent connection \cite{BanadosFerreira2010,BeltranJimenez2018}.

The most important property of $c_k$ is that, for every invertible matrix $S$,
\begin{equation}
 c_k(S\sfR S^{-1})=c_k(\sfR).
\end{equation}
If $r_i$ are the eigenvalues counted with algebraic multiplicity, then
\begin{equation}
 c_k=\sum_{i_1<\cdots<i_k}r_{i_1}\cdots r_{i_k}.
\end{equation}
This expression explicitly reveals the symmetry of $c_k$. Mathematically, this is precisely Vi\`ete's formula, which relates the roots of a polynomial to its coefficients and reflects the symmetry among the roots (here, the roots are the eigenvalues of the Ricci endomorphism). 

Within the class of scalar Lagrangians built from polynomial contractions of the Ricci tensor without its covariant derivatives, the $c_k$ form a natural set of invariant variables. We formulate the following \emph{Ricci spectral principle} as a model-building assumption: if the Ricci endomorphism is the chosen local curvature observable and no additional preferred direction is introduced, the local Lagrangian is taken to depend on its basis-independent spectral invariants. This assumption restricts the class of theories; it does not follow from general covariance alone and excludes independent Weyl-curvature invariants. 

Investigating ESRG theories with an arbitrary function \(F\) is evidently too broad in scope. Therefore, we specialize to the four-dimensional linear polynomial model
\begin{equation}
 F=-2\Lambda+R+\alpha c_2+\beta c_3+\gamma c_4.
 \label{equ:lin_model}
\end{equation}
Here ``linear'' refers to linearity in the $c_k$, rather than in curvature. 

 We note that a smooth local function $F(R^\mu{}_\nu)$ contains no derivatives of the curvature and generically produces fourth-order metric equations, as does the linear model above. Appendix~\ref{sec:app1} collects the invariant identities used below.

\section{Field equations and vacuum solutions} \label{sec:FE}
Metric variation of Eq.~\eqref{equ:equ_action} gives
\begin{equation}
 \Pi^\mu{}_{\alpha}R^\alpha{}_\nu-\tfrac12F\delta^\mu{}_\nu
 +\calD^\mu{}_\nu[\Pi]=\kappa T^\mu{}_\nu,
 \label{equ:evo1.1}
\end{equation}
where $\Pi^\mu{}_\nu=\partial F/\partial R^\nu{}_\mu$ and
\begin{align}
 \calD^\mu{}_\nu[\Pi]={}&\tfrac12\square\Pi^\mu{}_\nu
 +\tfrac12\delta^\mu{}_\nu\nabla_\alpha\nabla_\beta\Pi^{\alpha\beta}\nonumber\\
 &-\tfrac12\nabla_\alpha\bigl(\nabla^\mu\Pi_\nu{}^\alpha
 +\nabla_\nu\Pi^{\mu\alpha}\bigr).
 \label{equ:D}
\end{align}
Note that the derivative order in the last term is part of the definition and cannot be interchanged without curvature terms. Appendix~\ref{app:variation} derives this equation. 

Let $\sfPi$ denote the matrix with elements $\Pi^\mu{}_\nu$. The mathematical structure of $c_k$ allows $\sfPi$ to be expressed as a polynomial in $\sfR$ with scalar coefficients through the Newton transformations. Thus, $[\sfR,\sfPi]=0$. If $\sfR$ is diagonalizable, the same eigenbasis diagonalizes $\sfPi$. This property is particularly useful in our analysis of the vacuum solutions. 

Vacuum solutions can be classified as algebraic or dynamical. The former are simpler and include Ricci-flat metrics, Einstein metrics, and parallel-Ricci non-Einstein metrics, which will be discussed below. The latter are more complicated. We investigate an example in Sec.~\ref{sec:FLRW}, namely an FLRW vacuum solution. 

\subsection{Ricci-flat solutions}
For $\Lambda=0$, with the Ricci-flat ansatz $\sfR = 0$, we have  $F=0$, $\sfPi=I$, and the derivative terms vanish, so every Ricci-flat metric satisfies Eq.~\eqref{equ:evo1.1}. Consequently, the Schwarzschild and Kerr solutions remain exact vacuum solutions.  

\subsection{Einstein solutions}
For $R^\mu{}_\nu=\lambda\delta^\mu{}_\nu$, where $\lambda\in\RR$ is constant, the field equations reduce to an algebraic equation 
\begin{equation}
 P (\lambda):=2\Lambda-2\lambda+2\beta\lambda^3+\gamma\lambda^4=0.
 \label{equ:vacuum}
\end{equation}
Every Einstein metric satisfying this condition is a solution. Positive and negative roots label de Sitter and anti-de Sitter representatives when maximal symmetry is imposed. If $\beta=\gamma=0$, then $\lambda=\Lambda$. There are several notable properties of the Einstein vacua:

First, the coefficient $\alpha$ drops out of Eq.~\eqref{equ:vacuum}. However, it affects perturbations (see Sec.~\ref{sec:perturbation}); thus, the corresponding term is not topological (Appendix~\ref{app:vacua}). More generally, in even $D$, a term proportional to $c_{D/2}$ does not contribute to the Einstein-vacuum equation. 

Second, from the algebraic properties of Eq.~\eqref{equ:vacuum}, there are two kinds of boundaries in parameter space where the vacuum branches can change. One boundary is the locus satisfying $P(\lambda)=P'(\lambda)=0$, where roots collide and may subsequently enter or leave the real axis. Another boundary is $\gamma=0$, where the degree of the vacuum polynomial $P(\lambda)$ changes. On this boundary, a root can escape to infinity without a finite-root collision. At these boundaries, the vacuum structure changes, and the corresponding linearized theory may be critical or strongly coupled; see the perturbative stability analysis in Sec.~\ref{sec:perturbation}. 

Third, for $\Lambda\ne0$, a triple-root point at $\lambda=2\Lambda$ occurs when 
\begin{equation}
 (\beta,\gamma)=\left(\frac{1}{4\Lambda^2},-\frac{1}{8\Lambda^3}\right). \label{equ:triple} 
\end{equation}
These parameter values solve $P(\lambda)=P'(\lambda)=P''(\lambda)=0$. At this point, the vacuum is ultracritical. The linear and quadratic perturbations disappear for a certain value of $\alpha$. 

\begin{figure*}[t]
 \centering
\includegraphics[width=0.49\textwidth]{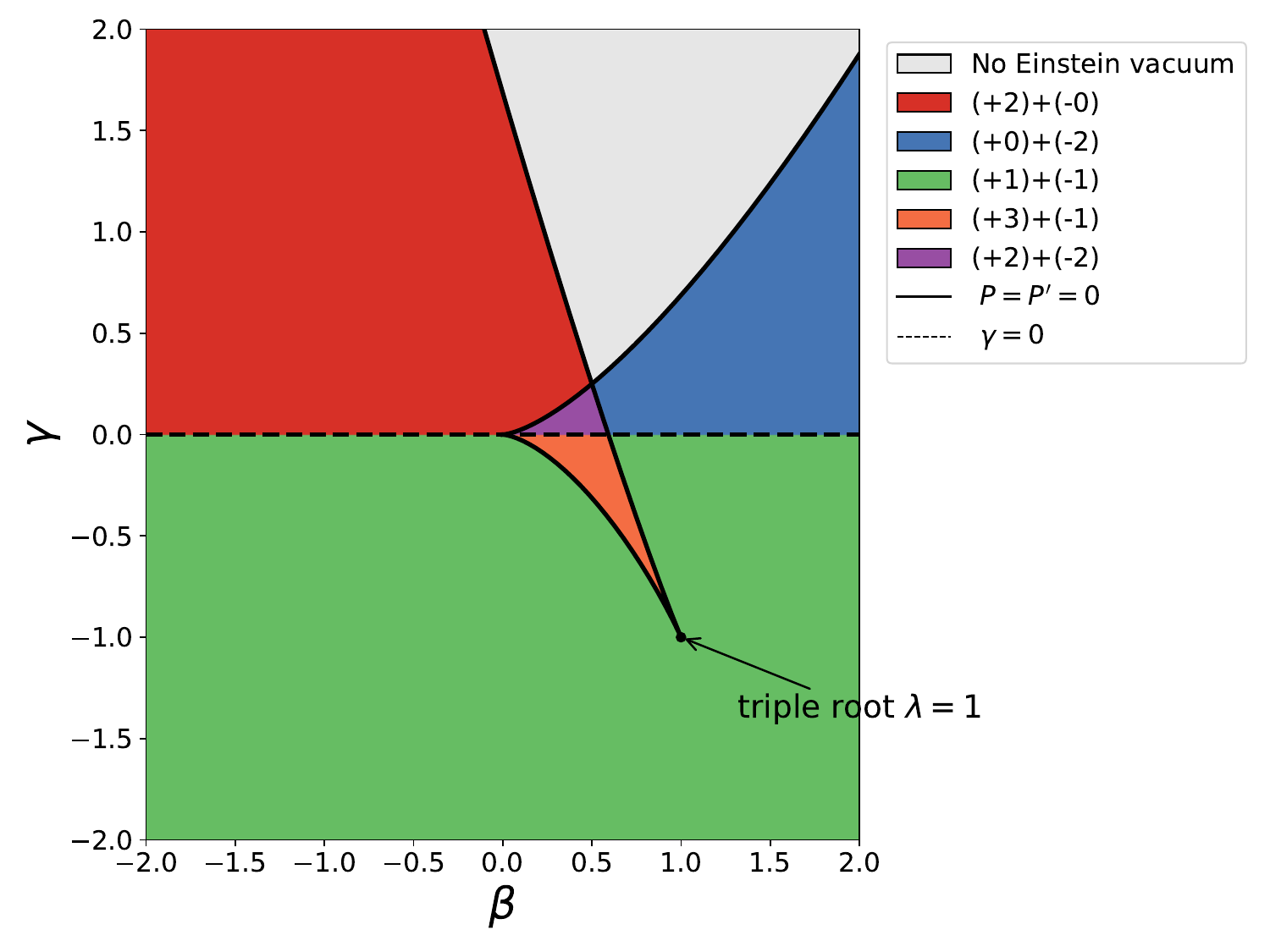}\hfill
 \includegraphics[width=0.47\textwidth]{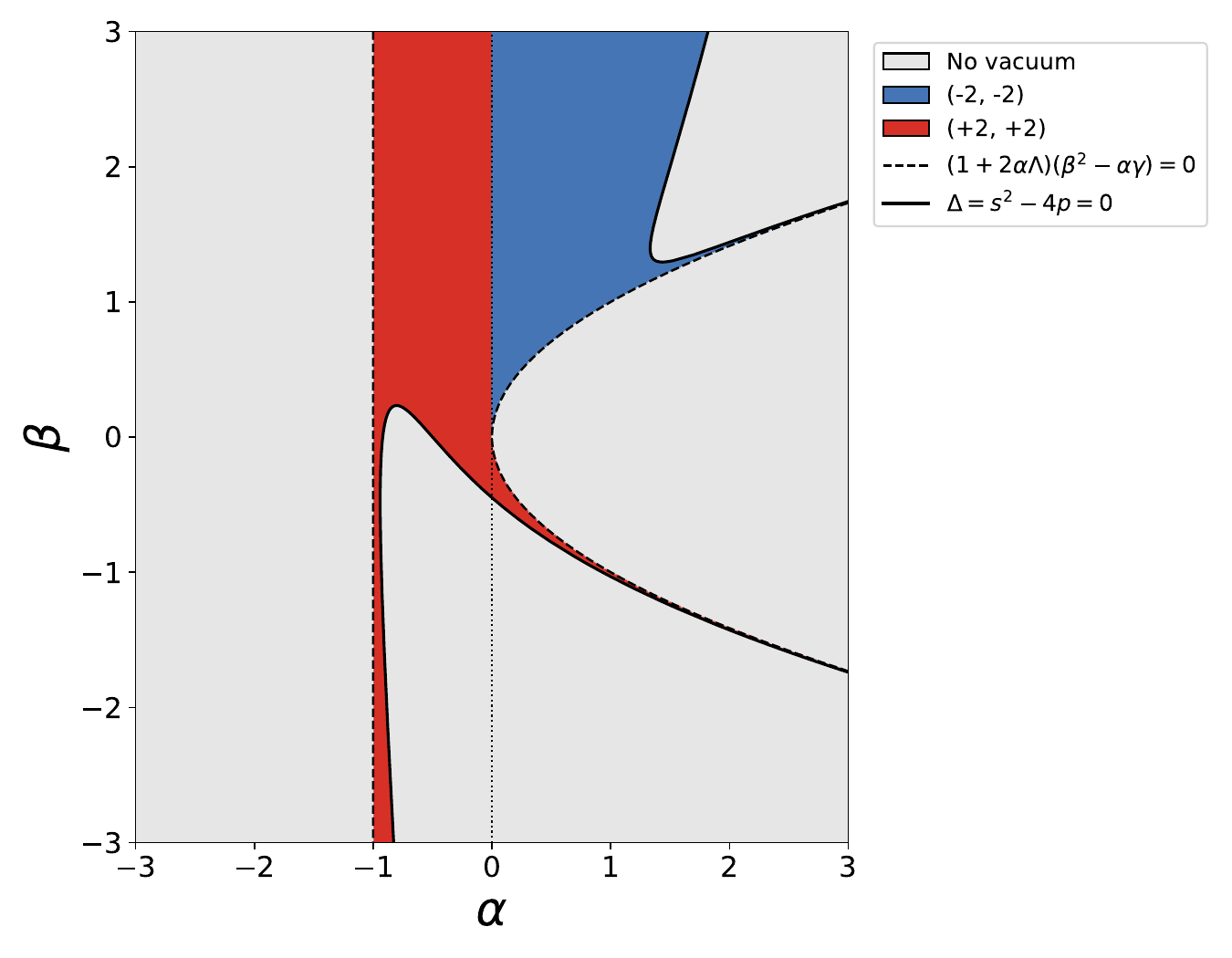}
 \caption{
 	Left: the vacuum phase diagram at $\Lambda=1/2$, 
 	where the vacua are classified by the sign of $\lambda$. For example, $(+2)+(-2)$ denotes a parameter region containing two positive and two negative Einstein-vacuum branches.     The solid curves satisfy $P=P'=0$; the dashed line is $\gamma=0$; and the labeled triple point satisfies $P(\lambda)=P'(\lambda)=P''(\lambda)=0$. 
 	Right: the $p>0 $ branch  of the non-Einstein $2+2$ family at $\Lambda=1/2$ and $\gamma=1$. The gray region has no real, distinct solution on this branch; note that it does not exclude negative-$p$ solutions. }
 \label{fig:dS_vaccum_phase}
\end{figure*}

The left panel of Fig.~\ref{fig:dS_vaccum_phase} shows the Einstein-vacuum phase diagram, in which the vacua are classified by the sign of $\lambda$. 

\subsection{Parallel-Ricci solutions}
For the parallel-Ricci ansatz 
\begin{equation}
 \nabla_\mu R^\rho{}_\sigma=0,
 \label{equ:prv_ansatz}
\end{equation}
the field equation becomes
\begin{equation}
 \sfPi\sfR=\tfrac12F I.
 \label{equ:prv}
\end{equation}
Parallel transport along a curve conjugates the Ricci endomorphism at its endpoints, so its eigenvalues are locally constant. We restrict here to real, diagonalizable Ricci endomorphisms with nondegenerate eigenspaces. In this class, the algebraic equations are 
\begin{align} 
\pi_i r_i =F/2, 
\end{align} 
where no summation over repeated indices is implied. The solutions can be classified according to the multiplicities of $r_i$. Taking the $2+2$ spectrum $(\lambda_1,\lambda_1,\lambda_2,\lambda_2)$ as an example, adding and subtracting the two independent eigenvalue equations yields
\begin{align}
 0={}&(\lambda_1-\lambda_2)
 \bigl[1+\alpha(\lambda_1+\lambda_2)+\beta\lambda_1\lambda_2\bigr],\nonumber\\
 0={}&2\Lambda-(\lambda_1+\lambda_2)
 +\beta\lambda_1\lambda_2(\lambda_1+\lambda_2)
 +\gamma\lambda_1^2\lambda_2^2.
 \label{equ:sp}
\end{align}
If $\lambda_1=\lambda_2$, one recovers the Einstein condition in Eq.~\eqref{equ:vacuum}. For distinct eigenvalues, the solution is given by
\begin{align}
	\lambda_{1,2}=\half\left(s\pm\sqrt{s^2-4p}\right) \label{equ:sp2l}
\end{align}
where, for $\alpha=0$, 
\begin{align}
	p =-  \frac{1}{\beta },\qquad 
	s = \Lambda + \frac{\gamma }{2 \beta^2 }
\end{align}
and for $\alpha\neq 0$, 
\begin{align}
	p = \pm   \sqrt{ \frac{1 + 2\alpha \Lambda }{\beta^2 - \alpha \gamma }}, \qquad 
	s =  - \frac{1+\beta p }{\alpha }
\end{align}

On the generic branch, real, distinct eigenvalues require $(\beta^2-\alpha\gamma)/(1+2\alpha\Lambda)\ge 0$ and $\Delta=s^2-4p>0$. 
For $p<0$ their signs are opposite. For $p>0$, both have the sign of $s$. The case $p=0$ includes a flat factor ($\lambda_2=0$) and is not covered by the nonzero-sign notation. The right panel of Fig.~\ref{fig:dS_vaccum_phase} shows the $p>0$ branch, where we use $(+2,+2)$ to label the vacuum with $\lambda_1,\lambda_2>0$ and $(-2,-2)$ to label the vacuum with $\lambda_1,\lambda_2<0$. The boundaries separating vacuum existence from nonexistence are $\Delta=0$ and $(\beta^2-\alpha\gamma)(1+2\alpha\Lambda)=0$. The half-line $\alpha=0$, $\beta>0$ is the boundary between $(-2,-2)$- and $(+2,+2)$-type vacua. 

Note that Eq.~\eqref{equ:prv} provides necessary conditions for parallel-Ricci vacua. A geometric construction is still required to establish existence. 
Products of two constant-curvature surfaces realize the $2+2$ spectrum. 
For a $(+2,+2)$ vacuum,
\begin{align}
 \dd s^2={}&-\left(1-\frac{r^2}{L_1^2}\right)\dd t^2
 +\frac{\dd r^2}{1-r^2/L_1^2}\nonumber\\
 &+L_2^2(\dd\theta^2+\sin^2\theta\,\dd\phi^2),
\end{align}
with $L_i=1/\sqrt{\lambda_i}$, describes a local patch of $\mathrm{dS}_2\times S^2$. For a $(-2,-2)$ vacuum, $L_i=1/\sqrt{|\lambda_i|}$ and
\begin{align}
 \dd s^2={}&-\left(1+\frac{r^2}{L_1^2}\right)\dd t^2
 +\frac{\dd r^2}{1+r^2/L_1^2}\nonumber\\
 &+L_2^2(\dd\chi^2+\sinh^2\chi\,\dd\phi^2)
\end{align}
gives $\mathrm{AdS}_2\times H^2$. Mixed signs are realized by $\mathrm{dS}_2\times H^2$ or $\mathrm{AdS}_2\times S^2$, depending on which eigenspace is Lorentzian. Related product geometries have a long history \cite{Nariai1951,Bertotti1959,Robinson1959}.

An interesting case is $s^2\gg|p|$, in which, according to Eq.~\eqref{equ:sp2l}, two roots with vastly different magnitudes necessarily emerge. Without loss of generality, one assumes $L_2\ll L_1$ in the $(+2,+2)$ branch, which suggests a compactification to two macroscopic dimensions.
The condition $s^2\gg|p|$ can be achieved, for instance,
by fixing $\beta$ and $\gamma$ while approaching the
critical value $\alpha=-1/(2\Lambda)$ from above.
Introducing $\epsilon=1+2\alpha\Lambda$, we have
$\epsilon\to 0^+$, which implies
$p\sim\sqrt{\epsilon}$ and $s \sim  2\Lambda$.
Consequently, $p/s^2\to 0$, and the condition
$s^2\gg|p|$ can be satisfied. 

\subsection{Nonparallel solutions}
Vacuum metrics with $\nabla\sfR\ne0$ are not determined by the preceding algebraic equations.  Without imposing some symmetries, these solutions are difficult to obtain. Time-dependent FLRW metrics provide an important setting in which to study this sector. Section~\ref{sec:FLRW} also includes matter sources.

\section{Linear perturbations and vacuum stability}\label{sec:perturbation} 
The detailed linear stability analysis depends on the background spacetime. For simplicity, we now restrict to maximally symmetric Einstein vacua, i.e., (A)dS vacua, which satisfy $\bar R_{\mu\nu\rho\sigma}=\frac{\lambda}{3}(\bar g_{\mu\rho}\bar g_{\nu\sigma}-\bar g_{\mu\sigma}\bar g_{\nu\rho})$, with $\lambda$ satisfying Eq.~\eqref{equ:vacuum}. 
Write $g_{\mu\nu}=\bar g_{\mu\nu}+\epsilon h_{\mu\nu}+O(\epsilon^2)$, and raise and lower perturbative indices with the background metric. Define
\begin{align}
 \mathcal R_{\mu\nu}&=\delta R_{\mu\nu},\nonumber\\
 \mathcal R&=\bar g^{\mu\nu}\mathcal R_{\mu\nu}-\lambda h,\nonumber\\
 \mathcal G^\mu{}_\nu
 &=\bar g^{\mu\alpha}\mathcal R_{\alpha\nu}-\lambda h^\mu{}_\nu
 -\tfrac12\mathcal R\delta^\mu{}_\nu,
\end{align}
where $\delta R_{\mu\nu}$ is given explicitly by Eq.~\eqref{equ:deltaRicci}. The background Bianchi identity implies $\bar\nabla_\mu\mathcal G^\mu{}_\nu=0$. 
With these definitions, the linear perturbation equation around an (A)dS vacuum in our model can be written as
\begin{align}
 q\mathcal G^\mu{}_\nu & 
 -\frac{B}{2}\left(\bar\square-\frac{2\lambda}{3}\right)\mathcal G^\mu{}_\nu\nonumber\\
 &+\frac{B}{2}\left(\delta^\mu{}_\nu\bar\square
 -\bar\nabla^\mu\bar\nabla_\nu
 +\frac{5\lambda}{3}\delta^\mu{}_\nu\right)\mathcal R = 0,  
 \label{equ:evo_lin}
\end{align}
where we have defined two coefficients that govern the linearized equation:
\begin{align}
	q(\lambda)&=1+3\alpha\lambda+3\beta\lambda^2+\gamma\lambda^3,\nonumber\\
	B(\lambda)&=\alpha+2\beta\lambda+\gamma\lambda^2.
	\label{equ:qB}
\end{align}
Appendix~\ref{app:linear} derives this expression directly and compares it using the equivalent quadratic action, following the general higher-curvature analysis of Refs.~\cite{HindawiOvrutWaldram1996a,HindawiOvrutWaldram1996b,GulluSismanTekin2011,Tekin2016}. A conserved matter perturbation appears as $\kappa\delta T^\mu{}_\nu$ on the right-hand side. In comoving perfect-fluid variables, $\delta T^\mu{}_\nu=\diag(-\delta\rho,\delta p,\delta p,\delta p)$ for a purely isotropic source.

\subsection{Generic scalar and tensor modes}
For $B\ne0$, we define $r(\lambda)=q(\lambda)/B(\lambda)$. The trace of Eq.~\eqref{equ:evo_lin} yields
\begin{equation}
 (\bar\square-m_0^2)\mathcal R=0,
\end{equation}
with 
\begin{align} 
	m_0^2
= \half (r - 3 \lambda )
\label{equ:m0}
\end{align} 
For the transverse-traceless metric perturbation,
$\bar\nabla^\mu h^{\mathrm{TT}}_{\mu\nu}=0$ and $h^{\mathrm{TT}}=0$, the equation factorizes as
\begin{equation}
 \left(\bar\square-\frac{2\lambda}{3}\right)
 \left(\bar\square-\frac{2\lambda}{3}-M_2^2\right)
 h^{\mathrm{TT}}_{\mu\nu}=0,
 \label{equ:TT}
\end{equation}
with
\begin{equation}
 M_2^2=2r.
 \label{equ:M2}
\end{equation}
For $r\ne0$ (i.e., $q\ne0$), the two Klein--Gordon-type factors represent massless and massive spin-2 branches. Their propagator residues have opposite signs, so positive squared masses do not make the full local theory ghost free \cite{Stelle1978,HindawiOvrutWaldram1996a}. Nevertheless, at the classical level, linear stability requires $m_0^2$ and $M_2^2$ to satisfy certain conditions, as analyzed below:

For de Sitter space with $\lambda>0$, a negative $m_0^2$ produces a growing homogeneous scalar solution, whereas $m_0^2>0$ gives scalar decay on the expanding patch. Thus, stability of the scalar component requires $m_0^2>0$, i.e., $r>3\lambda$. For the tensor component, the situation is more complex: de Sitter unitarity additionally requires the Higuchi bound $M_2^2\ge2\lambda/3$ \cite{Higuchi1987,DeserWaldron2001}, i.e., $r\ge\lambda/3$. 
In addition, $r=3\lambda$ and $r=0$ (or $q=0$, $B\ne0$) are two marginal points that require separate treatment, as discussed in Sec.~\ref{subsec:IVB}. The classification is summarized in Table~\ref{tab:dS_stability}. 

\begin{table*}[t]
	\caption{
		Linearized stability classification on a de Sitter vacuum.
		Here
		$r=q/B$,
		$m_0^2=(r-3\lambda)/2$,
		and
		$M_2^2=2r$.
		The Higuchi bound refers to an ordinary positive-sign Fierz--Pauli massive spin-2 field.
		Boundary cases require separate analysis.
	}
	\label{tab:dS_stability}
	\begin{ruledtabular}
		\begin{tabular}{lll}
			Condition
			&
			Scalar sector
			&
			Massive spin-2 sector
			\\
			\hline
			$r>3\lambda$
			&
			Stable, $m_0^2>0$
			&
			$M_2^2>0$; Higuchi bound satisfied
			\\
			$\lambda/3\le  r<3\lambda$
			&
			Unstable, $m_0^2<0$
			&
			$M_2^2>0$; Higuchi bound satisfied
			\\
			$0<r<\lambda/3$
			&
			Unstable, $m_0^2<0$
			&
			$0<M_2^2<2\lambda/3$; Higuchi bound violated
			\\
			$r<0$
			&
			Unstable, $m_0^2<0$
			&
			$M_2^2<0$
			\\
		\end{tabular}
	\end{ruledtabular}
\end{table*}

The corresponding sign test is insufficient in anti-de Sitter space. With $L^2=-3/\lambda$, a scalar may have a negative squared mass and still satisfy the Breitenlohner--Freedman bound $m_0^2L^2\ge-9/4$ \cite{BreitenlohnerFreedman1982}. Tensor stability, admissible falloff, and unitarity likewise require a separate boundary-condition analysis \cite{LuPangPope2011,Maldacena2011}. We therefore do not retain a universal AdS stable/unstable classification based only on mass signs.  


\subsection{Degenerate cases}\label{subsec:IVB}
If $q=0$ and $B\ne0$, the scalar component has $m_0^2=-3\lambda/2$, which is tachyonic in de Sitter space. For the tensor component, $M_2^2=0$, so
\begin{equation}
 \left(\bar\square-\frac{2\lambda}{3}\right)^2h^{\mathrm{TT}}_{\mu\nu}=0.
\end{equation}
The repeated tensor operator admits generalized solutions beyond the ordinary massless branch. Their admissibility depends on the background and boundary conditions; logarithmic modes in critical AdS gravity are a relevant example \cite{LuPope2011,DeserEtAl2011}. 

If $B=0$ and $q\ne0$, Eq.~\eqref{equ:evo_lin} reduces to $q\mathcal G^\mu{}_\nu=0$. Its trace gives $\mathcal R=0$, and only the massless tensor equation remains at linear order. A positive massless kinetic sign requires $q>0$. This reduction on a particular background is not a proof that the nonlinear theory has no additional strongly coupled degrees of freedom; related Ricci-polynomial constructions illustrate the importance of specifying the background \cite{LiLiuLu2018}.

If $q=B=0$, the entire linearized vacuum equation vanishes. The quadratic action supplies no bulk propagator, and higher-order perturbations are needed to determine the dynamics.  The vacuum becomes ultracritical. 

Finally, if $r=q/B=3\lambda$, the scalar mode becomes massless. At the same time,  
\begin{equation}
 P'(\lambda)=-2\bigl(q-3\lambda B\bigr)=0.
 \label{equ:vacuumderivative}
\end{equation}
Hence, two vacua merge at $r=3\lambda$. The emergence of the massless mode can be understood by analogy with a harmonic oscillator: at the point where the vacua merge, the linear restoring force for a small perturbation vanishes; hence the corresponding mass term, which plays the role of the restoring force, also disappears. 

Note that when $q=B=0$, Eq.~\eqref{equ:vacuumderivative} can also be satisfied. Thus, vacuum merging also occurs. However, this case differs from $r=3\lambda$. In the vacuum diagram, $q=B=0$ gives a point on the boundary that moves with the free parameter $\alpha$, i.e., $\beta=\beta(\alpha)$ and $\gamma=\gamma(\alpha)$. When this point moves to the triple root, $\alpha=-1/(2\Lambda)$.   In this case, together with Eq.~\eqref{equ:triple}, one obtains
\begin{align}
	F=-2\Lambda\det\left(I-\frac{1}{2\Lambda}\sfR\right).
\end{align}
The model degenerates into a Ricci-determinant gravity.

\begin{table*}[t]
\caption{Linearized mode structure on a maximally symmetric vacuum. Mass formulas apply when their denominators are nonzero; additional massless or partially massless limits require separate analysis.}
\label{tab:perturbation_sum}
\begin{ruledtabular}
\begin{tabular}{ll}
Condition & Generic content or degeneracy\\
\hline
$q\ne0$, $B\ne0$ & Scalar, massless spin-2, and massive spin-2; opposite tensor residues\\
$q=0$, $B\ne0$ & Scalar and repeated massless tensor operator\\
$q\ne0$, $B=0$ & Einstein-like massless spin-2 at quadratic order\\
$q=B=0$ & Vanishing bulk quadratic field equation\\
\end{tabular}
\end{ruledtabular}
\end{table*}

\section{FLRW solutions}\label{sec:FLRW}
Consider the FLRW metric
\begin{equation}
 \dd s^2=-\dd t^2+a^2(t)\bigl[\dd r^2+f_K^2(r)\dd\Omega_2^2\bigr],
 \label{equ:FE}
\end{equation}
where
\begin{equation}
 f_K(r)=\begin{cases}
 \sin(\sqrt K r)/\sqrt K,&K>0,\\
 r,&K=0,\\
 \sinh(\sqrt{-K}r)/\sqrt{-K},&K<0.
 \end{cases}
\end{equation}
Write $H=\dot a/a$, $X=\dot H$, and $k=K/a^2$. The Ricci spectrum is $(A,C,C,C)$, with
\begin{equation}
 A=3(X+H^2),\qquad C=X+3H^2+2k,
\end{equation}
The temporal and spatial eigenvalues of $\sfPi$ are
\begin{align}
 P_t&=1+3\alpha C+3\beta C^2+\gamma C^3,\nonumber\\
 P_s&=1+\alpha(A+2C)+\beta(2AC+C^2)+\gamma AC^2.
\end{align}
The $00$ field equation is the generalized Friedmann equation
\begin{align}
 -AP_t+\tfrac12F
 &+\tfrac32H(\dot P_t+\dot P_s)\nonumber\\
 &+\tfrac32(X+3H^2)(P_t-P_s)=\kappa\rho,
 \label{equ:fe}
\end{align}
where $\rho=\sum_m\rho_m$. Independently conserved perfect fluids satisfy
\begin{equation}
 \dot\rho_m=-3H(1+w_m)\rho_m,
 \qquad p_m=w_m\rho_m.
 \label{equ:FE2}
\end{equation}
Together with $\dot k=-2Hk$ and $\dot H=X$, these equations give an autonomous system on branches where the coefficient of $\dot X$ is nonzero. Appendix~\ref{app:flrw} derives the Friedmann equation by varying the lapse and supplies explicit evolution coefficients. Note that a fluid with $w_m=-1$ can instead be absorbed into $\Lambda$.

At a fixed point with $H_*\ne0$ and $w_m\ne-1$ for all remaining fluids, the conservation equations imply $\rho_{m*}=0$ and $k_*=0$, while $X_*=0$. Equation~\eqref{equ:fe} then becomes
\begin{equation}
 P (\lambda_*)=0,\qquad \lambda_*=3H_*^2>0.
\end{equation}
These fixed points describe expanding or contracting de Sitter patches. The scale factor itself is not stationary there; this is why it has been replaced by $k$ in the phase-space variables. The case $H_*=0$ lies on a singular branch of the reduced evolution equation and must be treated using the undivided field equations. For example, a static solution must also satisfy the spatial equation; the $00$ constraint alone is insufficient.

On the expanding, spatially flat vacuum branch with $B(\lambda_*)\ne0$, homogeneous perturbations satisfy
\begin{equation}
 \delta\ddot H+3H_*\delta\dot H+m_0^2\delta H=0.
 \label{equ:homogeneous}
\end{equation}
The corresponding eigenvalues are
\begin{equation}
 s_\pm=\frac{-3H_*\pm\sqrt{9H_*^2-4m_0^2}}{2}.
\end{equation}
Thus $m_0^2>0$ gives a homogeneous attractor and $m_0^2<0$ gives a saddle. For matter satisfying the null energy condition, with $w_m>-1$, the additional fluid eigenvalues $-3H_*(1+w_m)$ correspond to decaying modes. If, for a given set of parameters, the system possesses both a saddle point and an attractor, the vacuum may evolve from the vicinity of the saddle point toward the attractor. Near the saddle point, a suitable trajectory can exhibit a slow-roll phase~\cite{Starobinsky1980,Baumann2009}. 
To verify this possibility, we developed a numerical simulation and carried out the corresponding analysis.

We consider a Universe with dust and radiation. The illustrative parameters are given in Table~\ref{tab:simulation_parameters}.
The two positive roots are $\lambda_\ell\approx0.1001$
 and $\lambda_h\approx3.761$,
 which correspond respectively to $H_\ell=0.1826$
 and $H_h=1.120$, 
 with $m_{0\ell}^2\approx 2.38>0$ and $m_{0h}^2=-1.52<0$. The lower-curvature solution is a homogeneous attractor and the higher-curvature solution is a saddle. 
\begin{table}[h]
	\caption{Parameter values adopted in the numerical simulation; $\kappa=1$ in these units.}
	\label{tab:simulation_parameters}
	\begin{ruledtabular}
		\begin{tabular}{ccccccc}
			$\Lambda$ & $\alpha$ & $\beta$ & $\gamma$
			& $\rho_{m0}$ & $\rho_{r0}$ & $K$ \\
			\hline
			$0.1$ & $0.2$ & $0.05$ & $0.01$
			& $0.2$ & $0.4$ & $0$
		\end{tabular}
	\end{ruledtabular}
\end{table}

\begin{figure*}[t]
 \centering
 \includegraphics[width=0.49\textwidth]{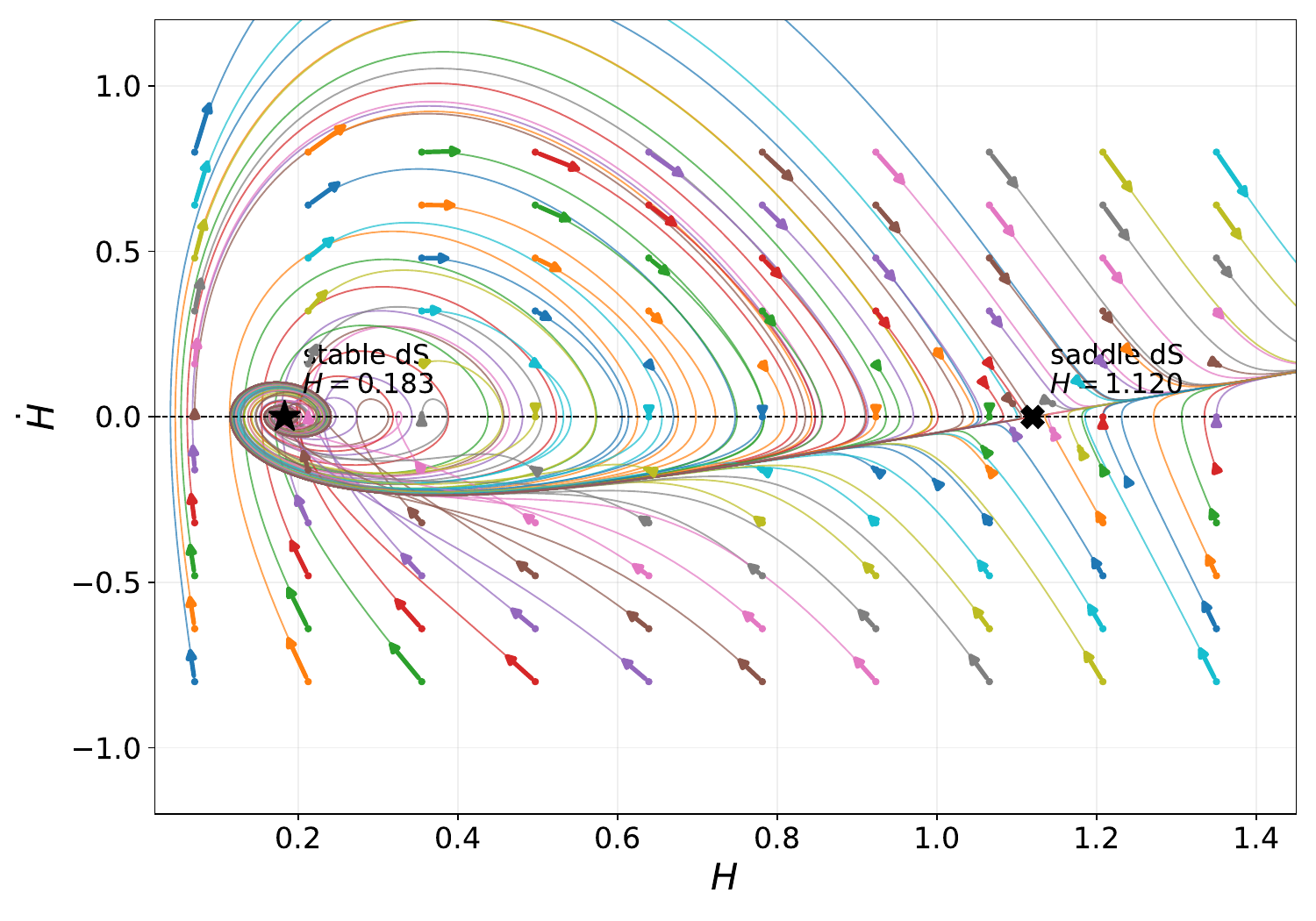}\hfill
 \includegraphics[width=0.45\textwidth]{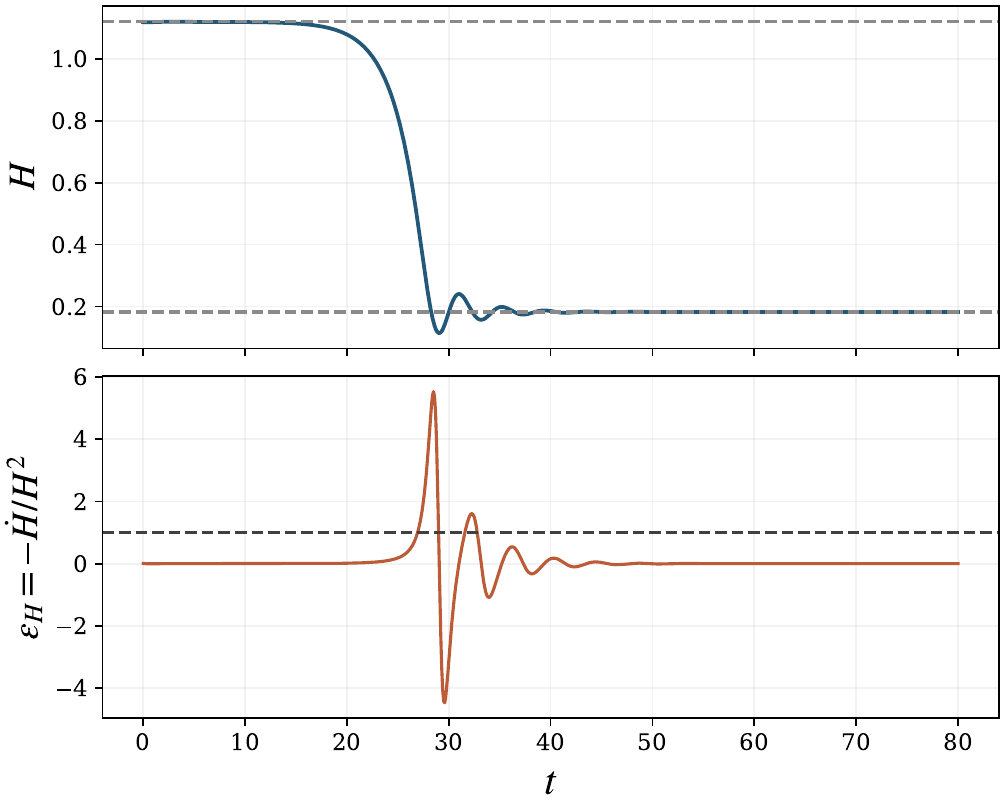}
 \caption{Left: the original projected FLRW trajectories with initial dust and radiation densities $(\rho_{m0},\rho_{r0})=(0.2,0.4)$, $K=0$, and the parameters stated in the text. Crossings in the $(H,\dot H)$ projection do not imply crossings in the full phase space. Arrows indicate the direction of evolution. Right: a representative dust--radiation trajectory near the upper de Sitter saddle, with $H(0)=H_h-0.0011$, $X(0)=-0.01$, $(\rho_m(0),\rho_r(0))=(0.2,0.4)$, and $K=0$. The dashed horizontal line in the lower panel marks $\epsilon_H=1$. The two panels use the same gravitational parameters and matter content.}
 \label{fig:flrw}
\end{figure*}

The left panel of Fig.~\ref{fig:flrw} shows the phase-space trajectories in the \((H,\dot H)\) plane, where the stable attractor and the unstable fixed point can be clearly identified. 
The right panel illustrates the evolution of $H$ and $\epsilon_H:=-\dot H/H^2$ along a representative dust--radiation trajectory. 
It starts at $H(0)=H_h-0.0011$, $X(0)=-0.01$\footnote{Setting $H(0) = H_h$ and $X(0) = 0$ exactly will lead to numerical instability. }, $\rho_m(0)=0.2$, $\rho_r(0)=0.4$, and $N(0)=0$, where $N=\ln[a/a(0)]$. Two fixed-step integrations with time steps $0.01$ and $0.005$ agree in $H$ to better than $1\times10^{-7}$ over $0\le t\le80$. For this initial condition, accelerated expansion first ends at approximately $t=26.94$, when $\epsilon_H$ reaches unity after $N\simeq28.51$. The duration depends on the initial displacement from the saddle. 
As can be seen, this trajectory indeed realizes our proposed cosmological scenario: the Universe evolves from an unstable high-\(H\) de Sitter state, undergoes a slow-roll phase, gradually exits the accelerated expansion phase, and eventually approaches a stable de Sitter attractor. 


\section{Discussion} \label{sec:discussion} 
The elementary symmetric polynomials provide a description of local Ricci-based metric actions. 
For the polynomial model considered here, we have briefly classified the vacuum solutions, investigated the stability of one particular class, and analyzed the cosmological dynamics in the FLRW background, for which the model supports a transition between two accelerated regimes within the homogeneous equations. A quantitative inflationary interpretation requires the duration of acceleration, scalar and tensor perturbation spectra, and a mechanism for producing a hot radiation bath \cite{MukhanovFeldmanBrandenberger1992,KofmanLindeStarobinsky1997}.  Furthermore, it is essential to confront the model with cosmic microwave background (CMB) observations from Planck and other missions \cite{Planck2018VI}. This requires first implementing the model in a cosmological Boltzmann solver such as \texttt{CLASS} \cite{Lesgourgues2011CLASS,BlasLesgourguesTram2011}.

Furthermore, the stability analysis of other algebraic vacua beyond the de Sitter class remains incomplete. Such an analysis is considerably more challenging due to the reduced symmetry of these backgrounds compared with de Sitter spacetime. From the perspective of stability, the Palatini formulation may offer a more suitable alternative. 
Moreover, the inclusion of nonlinear combinations of the curvature invariants, such as $c_2^2$, is entirely permissible, although such extensions may not introduce qualitatively new features. 
Nevertheless, due to limitations of space and the scope of the present work, we leave these investigations for future research.

\begin{acknowledgments}
	This work is supported by the National Key Research and Development Program (Nos. 2022YFA1602903 and 2023YFB3002502) and NSFC (Nos. 12595314, 12533007,  12503009, 12403007, 12573115
	and 12533008).  In this study, a cluster 'Zimo' is used with the SIMT accelerator made in China.  Xingpao Suo would like to thank Tingfei Li for helpful discussions.  
\end{acknowledgments}

\section*{CRediT authorship contribution statement}

\noindent
\textbf{Xingpao Suo:}
Conceptualization, Formal analysis, Writing -- original draft.

\noindent
\textbf{Dianxu Ma:}
Formal analysis.

\noindent
\textbf{Yuqiong Shan:}
Formal analysis.

\noindent
\textbf{Xingyan Suo:}
Software, Visualization.

\appendix
\section{Elementary symmetric polynomials and Newton transformations}
\label{sec:app1}
Define the generating polynomial $C(t)=\det(I+t\sfR)=\sum_{k=0}^{D}c_k t^k$. Expanding the characteristic polynomial after a shift and a rescaling gives
\begin{align}
 c_k(\lambda I+\sfR)&=\sum_{j=0}^k\binom{D-j}{k-j}\lambda^{k-j}c_j(\sfR),\label{equ:shift}\\
 c_k(a\sfR)&=a^k c_k(\sfR).
\end{align}
The rescaling identity is polynomial and also holds at $a=0$. Consequently, Euler's identity is
\begin{equation}
 R^\mu{}_\nu\frac{\partial c_k}{\partial R^\mu{}_\nu}=k c_k.
\end{equation}
Let $p_j=\tr(\sfR^j)$. In a neighborhood of $t=0$,
\begin{align}
 \frac{C'(t)}{C(t)}
 &=\tr\bigl[\sfR(I+t\sfR)^{-1}\bigr]\nonumber\\
 &=\sum_{m=0}^\infty(-1)^m p_{m+1}t^m.
\end{align}
Matching coefficients in $C'(t)=C(t)\sum_{m\ge0}(-1)^m p_{m+1}t^m$ gives the Newton identities
\begin{equation}
 k c_k=\sum_{j=1}^{k}(-1)^{j-1}c_{k-j}p_j.
\end{equation}
These are polynomial identities and require no diagonalizability assumption.

Varying the determinant gives
\begin{align}
 \delta C(t)
 &=t\,\tr\bigl[\operatorname{adj}(I+t\sfR)\,\delta\sfR\bigr]\nonumber\\
 &=\sum_{k=1}^{D}t^k\tr(T_{k-1}\delta\sfR),
\end{align}
where
\begin{equation}
 T_n=\sum_{j=0}^{n}(-1)^j c_{n-j}\sfR^j.
 \label{equ:newtonT}
\end{equation}
The adjugate identity is valid even when the matrix is singular; the intermediate inverse formula need only be used near $t=0$. In components,
\begin{equation}
 \delta c_k=(T_{k-1})^\nu{}_\mu\,\delta R^\mu{}_\nu,
 \qquad
 \frac{\partial c_k}{\partial R^\mu{}_\nu}=(T_{k-1})^\nu{}_\mu.
\end{equation}
The reversal of indices in the matrix derivative is essential. Therefore
\begin{equation}
 \sfPi=\sum_{k=1}^{D}F_{,c_k}T_{k-1},
\end{equation}
and, in the model of Eq.~\eqref{equ:lin_model},
\begin{align}
 \sfPi={}&I+\alpha(c_1I-\sfR)
 +\beta(c_2I-c_1\sfR+\sfR^2)\nonumber\\
 &+\gamma(c_3I-c_2\sfR+c_1\sfR^2-\sfR^3).
 \label{equ:Pi_explicit}
\end{align}

\section{Metric variation}
\label{app:variation}
Consider compactly supported variations, so that boundary terms can be omitted in deriving the bulk equations. With $\Pi^{\mu\nu}=g^{\nu\alpha}\Pi^\mu{}_\alpha$, symmetry follows from the Ricci-polynomial construction. For a variation of the inverse metric,
\begin{align}
 \delta\sfR^\mu{}_\nu
 &=R_{\alpha\nu}\delta g^{\mu\alpha}
   +g^{\mu\alpha}\delta R_{\alpha\nu},\nonumber\\
 \delta F&=\Pi^\nu{}_\mu R_{\alpha\nu}\delta g^{\mu\alpha}
             +\Pi^{\mu\nu}\delta R_{\mu\nu}.
\end{align}
For $h_{\mu\nu}=\delta g_{\mu\nu}$,
\begin{align}
 \delta R_{\mu\nu}
 =\tfrac12\bigl(&\nabla_\alpha\nabla_\mu h^\alpha{}_\nu
 +\nabla_\alpha\nabla_\nu h^\alpha{}_\mu\nonumber\\
 &-\square h_{\mu\nu}-\nabla_\mu\nabla_\nu h\bigr).
 \label{equ:deltaRicci}
\end{align}
Integrating twice by parts and using $\delta g_{\mu\nu}=-g_{\mu\alpha}g_{\nu\beta}\delta g^{\alpha\beta}$ yields
\begin{align}
 \frac{2\kappa}{\sqrt{-g}}\frac{\delta S_g}{\delta g^{\mu\nu}}
 ={}&\Pi_{(\mu}{}^\alpha R_{\nu)\alpha}-\tfrac12Fg_{\mu\nu}
 +\tfrac12\square\Pi_{\mu\nu}\nonumber\\
 &+\tfrac12g_{\mu\nu}\nabla_\alpha\nabla_\beta\Pi^{\alpha\beta}
 -\nabla_\alpha\nabla_{(\mu}\Pi_{\nu)}{}^\alpha.
\end{align}
Parentheses denote symmetrization with weight $1/2$. Since $\sfPi$ commutes with $\sfR$, the first term is symmetric without an explicit symmetrizer. The definition of the matter stress-energy tensor, $T_{\mu\nu}=-2(-g)^{-1/2}\delta S_m/\delta g^{\mu\nu}$, gives Eq.~\eqref{equ:evo1.1}. As checks, $F=R-2\Lambda$ reproduces the Einstein equation and $F=f(R)$ gives the derivative term $(g_{\mu\nu}\square-\nabla_\mu\nabla_\nu)f'(R)$.

\section{Algebraic vacuum equations and product geometries}
\label{app:vacua}
For a general linear invariant action $F=\sum_{k=0}^{D}a_k c_k$, an Einstein background obeys
\begin{equation}
 c_k=\binom Dk\lambda^k,\qquad
 \lambda\pi=\sum_{k=0}^{D}\frac{k}{D}a_k\binom Dk\lambda^k.
\end{equation}
Thus $\lambda\pi-F/2=0$ is
\begin{equation}
 \sum_{k=0}^{D}\left(\frac{k}{D}-\frac12\right)
 a_k\binom Dk\lambda^k=0.
\end{equation}
The coefficient of $a_{D/2}$ vanishes when $D$ is even. For $D=4$ this gives Eq.~\eqref{equ:vacuum}. Differentiating that polynomial gives the multiple-root relations in Eq.~\eqref{equ:fold}.

For the spectrum $(x,x,y,y)$, direct expansion gives
\begin{align}
 c_1&=2(x+y),&c_2&=x^2+4xy+y^2,\nonumber\\
 c_3&=2xy(x+y),&c_4&=x^2y^2,
\end{align}
so that
\begin{align}
 F={}&-2\Lambda+2(x+y)+\alpha(x^2+4xy+y^2)\nonumber\\
 &+2\beta xy(x+y)+\gamma x^2y^2,\nonumber\\
 \pi_x={}&1+\alpha(x+2y)+\beta(2xy+y^2)+\gamma xy^2,\nonumber\\
 \pi_y={}&1+\alpha(2x+y)+\beta(2xy+x^2)+\gamma x^2y.
\end{align}
The multiplicity factors imply $\pi_x=\tfrac12\partial F/\partial x$, rather than $\partial F/\partial x$. Subtracting $y\pi_y$ from $x\pi_x$ gives $(x-y)[1+\alpha(x+y)+\beta xy]$. Adding and subtracting $F$ gives the second equation in~\eqref{equ:sp}. Eliminating $s$ for $\alpha\ne0$ yields
\begin{equation}
 (\beta^2-\alpha\gamma)p^2=1+2\alpha\Lambda.
\end{equation}
This proves the generic solution and helps identify the exceptional cases discussed in the main text.

If $U$ denotes parallel transport along any specified curve, $\nabla\sfR=0$ implies $\sfR(x)=U\sfR(x_0)U^{-1}$. This statement does not assert that $U$ is path independent. For a direct product of two two-dimensional constant-curvature metrics, the product connection preserves each factor, and the Ricci tensor equals $xg_{(1)}\oplus yg_{(2)}$. Hence its covariant derivative vanishes and the algebraic conditions above suffice for an exact solution. We do not assume that all Lorentzian parallel-Ricci metrics admit this product form.

\section{Linearization about a maximally symmetric vacuum}
\label{app:linear}

This appendix derives the linearized metric equations directly from the
field equations and checks the result by an independent expansion of the
action.  We use signature $(-,+,+,+)$ and the curvature convention
\begin{equation}
 \bar R_{\mu\nu\rho\sigma}
 =\frac{\lambda}{3}
 (\bar g_{\mu\rho}\bar g_{\nu\sigma}
 -\bar g_{\mu\sigma}\bar g_{\nu\rho}),
 \qquad \bar R_{\mu\nu}=\lambda\bar g_{\mu\nu}.
 \label{eq:app_maxsym}
\end{equation}
Maximal symmetry is essential for the simple wave operators below;
$\bar R_{\mu\nu}=\lambda\bar g_{\mu\nu}$ alone does not imply
Eq.~\eqref{eq:app_maxsym}.  The background is assumed to solve
$P (\lambda)=0$.

\subsection{First variations and the Ricci Hessian}

Set $g_{\mu\nu}=\bar g_{\mu\nu}+\epsilon h_{\mu\nu}+O(\epsilon^2)$,
with all perturbative indices raised using $\bar g$.  The coefficient of
$\epsilon$ in the covariant Ricci tensor is
\begin{align}
 \mathcal R_{\mu\nu}={}&\frac12\bigl(
 \bar\nabla_\rho\bar\nabla_\mu h^\rho{}_{\nu}
 +\bar\nabla_\rho\bar\nabla_\nu h^\rho{}_{\mu}
 -\bar\square h_{\mu\nu}
 -\bar\nabla_\mu\bar\nabla_\nu h\bigr),
 \label{eq:app_Ricci_variation}\\
 X^\mu{}_{\nu}:={}&\delta R^\mu{}_{\nu}
 =\bar g^{\mu\rho}\mathcal R_{\rho\nu}-\lambda h^\mu{}_{\nu},
 \qquad \mathcal R=\operatorname{tr}X.
 \label{eq:app_X}
\end{align}
Writing the first two terms separately avoids any ambiguity in the
normalization of symmetrization.  Define the variation of the
cosmological Einstein tensor by
\begin{equation}
 \mathcal G_{\mu\nu}
 =\mathcal R_{\mu\nu}-\lambda h_{\mu\nu}
 -\frac12\bar g_{\mu\nu}\mathcal R.
 \label{eq:app_G}
\end{equation}
It obeys $\bar\nabla^\mu\mathcal G_{\mu\nu}=0$ and
$\bar g^{\mu\nu}\mathcal G_{\mu\nu}=-\mathcal R$.

For a small matrix $Y$, the exact translation identity for elementary
symmetric polynomials gives
\begin{align}
 c_k(\lambda I+Y)={}&\binom4k\lambda^k\nonumber\\
 &+\binom3{k-1}\lambda^{k-1}\operatorname{tr}Y\nonumber\\
 &+\frac12\binom2{k-2}\lambda^{k-2}\nonumber\\
 &\quad\times\bigl[(\operatorname{tr}Y)^2-\operatorname{tr}(Y^2)\bigr]
 +O(Y^3),
 \label{eq:app_ck_expand}
\end{align}
where terms outside the binomial range are omitted.  Consequently,
\begin{align}
 F(\lambda I+Y)={}&\bar F+q\operatorname{tr}Y\nonumber\\
 &+\frac B2\bigl[(\operatorname{tr}Y)^2-\operatorname{tr}(Y^2)\bigr]
 +O(Y^3),\\
 q={}&1+3\alpha\lambda+3\beta\lambda^2+\gamma\lambda^3,\nonumber\\
 B={}&\alpha+2\beta\lambda+\gamma\lambda^2.
 \label{eq:app_qB}
\end{align}
Thus $\bar\Pi^\mu{}_{\nu}=q\delta^\mu{}_{\nu}$ and
\begin{equation}
 \delta\Pi^\mu{}_{\nu}
 =B(\mathcal R\delta^\mu{}_{\nu}-X^\mu{}_{\nu})
 =B\left(\frac12\mathcal R\delta^\mu{}_{\nu}
 -\mathcal G^\mu{}_{\nu}\right).
 \label{eq:app_dPi}
\end{equation}

\subsection{Direct variation of the field equation}

The algebraic part of the mixed field equation varies as
\begin{equation}
 \delta\left(\Pi^\mu{}_{\rho}R^\rho{}_{\nu}
 -\frac12F\delta^\mu{}_{\nu}\right)
 =q\mathcal G^\mu{}_{\nu}
 +\lambda B\left(\frac12\mathcal R\delta^\mu{}_{\nu}
 -\mathcal G^\mu{}_{\nu}\right).
 \label{eq:app_algebraic}
\end{equation}
Because the identity endomorphism is covariantly constant for every
metric, the derivative terms acting on $qI$ vanish identically, including
their first variations.  The remaining derivative term is the
background operator acting on Eq.~\eqref{eq:app_dPi}.  In particular,
maximal symmetry and the linearized Bianchi identity give
\begin{equation}
 \bar\nabla_\rho\bar\nabla_{(\mu}
 \mathcal G_{\nu)}{}^\rho
 =\frac{4\lambda}{3}\mathcal G_{\mu\nu}
 +\frac{\lambda}{3}\bar g_{\mu\nu}\mathcal R,
 \label{eq:app_commutator}
\end{equation}
where parentheses have weight $1/2$.  To see the curvature terms
explicitly, commute the first two derivatives, use
$\bar\nabla_\rho\mathcal G^\rho{}_{\nu}=0$, and apply
\begin{equation}
 [\bar\nabla_\rho,\bar\nabla_\mu]
 \mathcal G^\rho{}_{\nu}
 =\bar R^\rho{}_{\sigma\rho\mu}\mathcal G^\sigma{}_{\nu}
 -\bar R^\sigma{}_{\nu\rho\mu}\mathcal G^\rho{}_{\sigma}.
\end{equation}
It follows that
\begin{align}
 \delta\mathcal D_{\mu\nu}={}&
 -\frac B2\bar\square\mathcal G_{\mu\nu}
 +\frac{4\lambda B}{3}\mathcal G_{\mu\nu}
 +\frac B2(\bar g_{\mu\nu}\bar\square
 -\bar\nabla_\mu\bar\nabla_\nu)\mathcal R\nonumber\\
 &+\frac{\lambda B}{3}\bar g_{\mu\nu}\mathcal R.
\end{align}
Combining this result with Eq.~\eqref{eq:app_algebraic} yields
\begin{align}
 0={}&\left[q-\frac B2\left(\bar\square-\frac{2\lambda}{3}\right)
 \right]\mathcal G_{\mu\nu}\nonumber\\
 &+\frac B2\left(\bar g_{\mu\nu}\bar\square
 -\bar\nabla_\mu\bar\nabla_\nu
 +\frac{5\lambda}{3}\bar g_{\mu\nu}\right)\mathcal R.
 \label{eq:app_linear_final}
\end{align}
The terms proportional to $\lambda B$ in the algebraic and derivative
variations must both be retained; their combination leaves the
coefficient $q$ in the form displayed here.

\subsection{Independent check using an equivalent quadratic action}

The expansion above also gives a Lagrangian with the same background
and linearized field equations:
\begin{align}
 F_{\mathrm{eq}}={}&C+(q-3\lambda B)R
 +\frac B2(R^2-R_{\mu\nu}R^{\mu\nu}),\\
 C={}&\bar F-4\lambda q+6\lambda^2B
 =-2\lambda(q-3\lambda B).
 \label{eq:app_equiv_quadratic}
\end{align}
The last equality uses $\bar F=2\lambda q$, which is exactly the
background field equation.  The difference $F-F_{\mathrm{eq}}$ begins
at cubic order in $R^\mu{}_{\nu}-\lambda\delta^\mu{}_{\nu}$ and hence
cannot affect the quadratic metric action.  Substitution of the
coefficients in Eq.~\eqref{eq:app_equiv_quadratic} into the general
quadratic-gravity linearization reproduces
Eq.~\eqref{eq:app_linear_final}~\cite{LuPope2011}.

\subsection{Masses, critical loci, and the meaning of stability}

Tracing Eq.~\eqref{eq:app_linear_final} gives
\begin{equation}
 [2B\bar\square-q+3\lambda B]\mathcal R=0.
 \label{eq:app_trace}
\end{equation}
For $B\ne0$, the scalar mass is therefore
\begin{equation}
 m_0^2=\frac{q-3\lambda B}{2B}
 =\frac{1-3\beta\lambda^2-2\gamma\lambda^3}{2B}.
 \label{eq:app_scalar_mass}
\end{equation}
In the transverse-traceless spin-2 sector, $\mathcal R=0$ and
$\mathcal G_{\mu\nu}=-\tfrac12\mathcal L h^{TT}_{\mu\nu}$, where
$\mathcal L=\bar\square-2\lambda/3$.  Thus
\begin{equation}
 \mathcal L(\mathcal L-M_2^2)h^{TT}_{\mu\nu}=0,
 \qquad M_2^2=\frac{2q}{B}.
 \label{eq:app_spin2_mass}
\end{equation}
This establishes the generic scalar, massless spin-2, and massive
spin-2 branches.  The spin-2 kinetic polynomial is proportional to
$\mathcal L(q-B\mathcal L/2)$.  When $qB\ne0$, its inverse decomposes as
\begin{equation}
 \frac{1}{\mathcal L(q-B\mathcal L/2)}
 =\frac1q\left[\frac1{\mathcal L}
 -\frac1{\mathcal L-2q/B}\right].
 \label{eq:app_opposite_residues}
\end{equation}
The relative minus sign shows that the two spin-2 poles cannot both
have positive residues in the conventional local quantization.
Positive squared masses therefore do not establish freedom from
ghosts or nonlinear stability.

For an expanding de Sitter patch, $H^2=\lambda/3>0$, a spatially
homogeneous scalar perturbation obeys
\begin{equation}
 \ddot{\mathcal R}+3H\dot{\mathcal R}+m_0^2\mathcal R=0,
 \qquad
 s_\pm=\frac{-3H\pm\sqrt{9H^2-4m_0^2}}2.
 \label{eq:app_dS_growth}
\end{equation}
Thus $m_0^2>0$ gives decay, $m_0^2<0$ gives one growing scalar mode,
and $m_0^2=0$ is marginal at linear order.  The usual Higuchi bound
for a massive spin-2 field with positive kinetic sign is
$M_2^2\ge2\lambda/3$~\cite{Higuchi1987}; it is a norm condition
separate from the absence of exponentially growing solutions.  In the
present model, $m_0^2>0$ implies $q/B>3\lambda$ and hence
$M_2^2>6\lambda$, but Eq.~\eqref{eq:app_opposite_residues} still
precludes a simultaneously positive-residue pair of spin-2 poles.

For anti-de Sitter space, let $\lambda=-3/\ell^2$.  The scalar
Breitenlohner--Freedman bound with the usual admissible boundary
conditions is
$m_0^2\ell^2\ge-9/4$, or $m_0^2\ge3\lambda/4$
\cite{BreitenlohnerFreedman1982}.  Negative $m_0^2$ alone is therefore
not an instability criterion.  For the massive spin-2 equation above,
the lowest-energy parameter is
$E_0=\tfrac32+\sqrt{\tfrac94+M_2^2\ell^2}$.
Reality requires $M_2^2\ge3\lambda/4$, whereas the unitary
spin-2 representation bound $E_0\ge3$ requires $M_2^2\ge0$
\cite{LuPangPope2011}.  Boundary conditions and the kinetic residues
must be assessed in addition to these mass inequalities; the de Sitter
sign classification cannot simply be transferred to anti-de Sitter
space.

Finally, three distinct degeneracies should be distinguished.
For $q=0$ and $B\ne0$, the spin-2 equation is
$\mathcal L^2h^{TT}_{\mu\nu}=0$ and admits generalized solutions in
addition to ordinary massless modes, including logarithmic modes for
appropriate AdS coordinates and boundary conditions.  The scalar
remains present with $m_0^2=-3\lambda/2$.
For $B=0$ and $q\ne0$, the entire linearized equation reduces to
$q\mathcal G_{\mu\nu}=0$, so only the Einstein graviton propagates at
this order; this does not prove nonlinear equivalence to GR.
For $q=B=0$, the quadratic metric action is degenerate and the
linearized field equation vanishes identically.  Higher-order terms
are then required to determine the dynamics; the linear analysis
alone gives neither a propagator nor a quantitative strong-coupling
scale.
The vacuum polynomial satisfies the exact identity
\begin{equation}
 P'(\lambda)=-2(q-3\lambda B).
 \label{eq:app_Pprime}
\end{equation}
Hence a finite multiple vacuum root with $B\ne0$ has a massless
scalar, but need not have $q=0$.  Conversely, $q=0$ alone does not
make the vacuum root multiple.  The fully degenerate case $q=B=0$
is necessarily a multiple root.  These are algebraic and perturbative
degeneracies; establishing a thermodynamic phase transition requires
additional information.

\section{Lapse variation and homogeneous evolution}
\label{app:flrw}
Retain the lapse $n(t)$ until after variation:
\begin{equation}
 \dd s^2=-n^2\dd t^2+a^2\dd\Sigma_K^2,\qquad
 H_n=\frac{\dot a}{na}.
\end{equation}
The Ricci eigenvalues are $A_n=3(n^{-1}\dot H_n+H_n^2)$ and $C_n=n^{-1}\dot H_n+3H_n^2+2k$. At $n=1$, write $\delta n=\eta$ and hold $a$ fixed. Then
\begin{equation}
 \delta A=-2A\eta-3H\dot\eta,\qquad
 \delta C=-2(C-2k)\eta-H\dot\eta.
\end{equation}
On the spectrum $(A,C,C,C)$,
\begin{align}
 F={}&-2\Lambda+A+3C+3\alpha C(A+C)\nonumber\\
 &+\beta C^2(3A+C)+\gamma AC^3,
\end{align}
and $F_{,A}=P_t$, $F_{,C}=3P_s$. Varying $\int\dd t\,na^3F/(2\kappa)$ and integrating the $\dot\eta$ term by parts gives
\begin{align}
 \kappa\rho=\tfrac12\bigl[&F-2AP_t-6(C-2k)P_s\nonumber\\
 &+3(X+3H^2)(P_t+P_s)
 +3H(\dot P_t+\dot P_s)\bigr].
\end{align}
Here $\delta S_m=-\int\dd t\,a^3\rho\eta$. Using $C-2k=X+3H^2$ immediately reproduces Eq.~\eqref{equ:fe}. For $\alpha=\beta=\gamma=0$ it reduces to $3(H^2+k)-\Lambda=\kappa\rho$.

For explicit evolution, define
\begin{equation}
 U=\alpha+2\beta C+\gamma C^2,\qquad
 W=\alpha+\beta(A+C)+\gamma AC.
\end{equation}
Direct differentiation gives
\begin{align}
 P_t-P_s&=(C-A)U,\nonumber\\
 \dot P_t+\dot P_s&=U\dot A+(3U+2W)\dot C,\nonumber\\
 \dot A&=3\dot X+6HX,\nonumber\\
 \dot C&=\dot X+6HX-4Hk.
\end{align}
Consequently the constraint becomes
\begin{equation}
 \mathcal A\dot X+\mathcal B X+\mathcal C=0,
 \label{equ:autoconstraint}
\end{equation}
where
\begin{align}
 \mathcal A={}&3H(3U+W),\nonumber\\
 \mathcal B={}&18H^2(2U+W)-3(C-2k)U,\nonumber\\
 \mathcal C={}&-AP_t+\tfrac12F-\kappa\rho\nonumber\\
 &+k\bigl[3(C-2k)U-6H^2(3U+2W)\bigr].
 \label{equ:autocoeffs}
\end{align}
When $\mathcal A\ne0$, one may evolve
\begin{equation}
 \dot H=X,\qquad
 \dot X=-\frac{\mathcal B X+\mathcal C}{\mathcal A},
 \qquad \dot k=-2Hk,
\end{equation}
together with Eq.~\eqref{equ:FE2}. The division is not valid at $H=0$ or $3U+W=0$. Those loci require the original equations and a separate constraint analysis.

For completeness, let $\mathcal E^\mu{}_\nu$ denote the left side of the gravitational field equation and set $\mathcal E^0{}_0=-E_\rho$, $\mathcal E^i{}_j=E_p\delta^i{}_j$. Diffeomorphism invariance gives $\dot E_\rho+3H(E_\rho+E_p)=0$. Matter conservation therefore allows the spatial field equation to be inferred from the $00$ equation on intervals where $H\ne0$. It supplies no such inference for an identically static solution.

At a flat de Sitter fixed point, $A=C=\lambda_*=3H_*^2$, $U=W=B(\lambda_*)$, and $\mathcal A=12H_*B$. Linearizing Eq.~\eqref{equ:autoconstraint} in vacuum gives
\begin{equation}
 12H_*B\,\delta\ddot H+36H_*^2B\,\delta\dot H
 +6H_*[q-3\lambda_*B]\delta H=0.
\end{equation}
Division by $12H_*B$ yields Eq.~\eqref{equ:homogeneous} and independently verifies the scalar mass obtained from the covariant perturbation equations. The evolution equation for $N$ is simply $\dot N=H$; $N$ is reconstructed after solving for $H$, rather than included when classifying fixed points.

\bibliographystyle{apsrev4-2}
\bibliography{apssamp}

@article{SotiriouFaraoni2010,
  author = {Sotiriou, Thomas P. and Faraoni, Valerio},
  title = {{f(R)} Theories of Gravity},
  journal = {Rev. Mod. Phys.},
  volume = {82}, pages = {451--497}, year = {2010},
  doi = {10.1103/RevModPhys.82.451},
  eprint = {0805.1726}, archivePrefix = {arXiv}, primaryClass = {gr-qc}
}

@article{DeFeliceTsujikawa2010,
  author = {De Felice, Antonio and Tsujikawa, Shinji},
  title = {{f(R)} Theories},
  journal = {Living Rev. Relativ.},
  volume = {13}, pages = {3}, year = {2010},
  doi = {10.12942/lrr-2010-3},
  eprint = {1002.4928}, archivePrefix = {arXiv}, primaryClass = {gr-qc}
}

@article{HindawiOvrutWaldram1996a,
  author = {Hindawi, Ahmed and Ovrut, Burt A. and Waldram, Daniel},
  title = {Consistent Spin-Two Coupling and Quadratic Gravitation},
  journal = {Phys. Rev. D},
  volume = {53}, pages = {5583--5596}, year = {1996},
  doi = {10.1103/PhysRevD.53.5583},
  eprint = {hep-th/9509142}, archivePrefix = {arXiv}
}

@article{HindawiOvrutWaldram1996b,
  author = {Hindawi, Ahmed and Ovrut, Burt A. and Waldram, Daniel},
  title = {Non-Trivial Vacua in Higher-Derivative Gravitation},
  journal = {Phys. Rev. D},
  volume = {53}, pages = {5597--5608}, year = {1996},
  doi = {10.1103/PhysRevD.53.5597},
  eprint = {hep-th/9509147}, archivePrefix = {arXiv}
}

@article{DeserGibbons1998,
  author = {Deser, S. and Gibbons, G. W.},
  title = {{Born--Infeld--Einstein} Actions?},
  journal = {Class. Quantum Grav.},
  volume = {15}, pages = {L35--L39}, year = {1998},
  doi = {10.1088/0264-9381/15/5/001},
  eprint = {hep-th/9803049}, archivePrefix = {arXiv}
}

@article{Olmo2011,
  author = {Olmo, Gonzalo J.},
  title = {{Palatini} Approach to Modified Gravity: {f(R)} Theories and Beyond},
  journal = {Int. J. Mod. Phys. D},
  volume = {20}, pages = {413--462}, year = {2011},
  doi = {10.1142/S0218271811018925},
  eprint = {1101.3864}, archivePrefix = {arXiv}, primaryClass = {gr-qc}
}

@article{LuPope2011,
  author = {L{\"u}, H. and Pope, C. N.},
  title = {Critical Gravity in Four Dimensions},
  journal = {Phys. Rev. Lett.},
  volume = {106}, pages = {181302}, year = {2011},
  doi = {10.1103/PhysRevLett.106.181302},
  eprint = {1101.1971}, archivePrefix = {arXiv}, primaryClass = {hep-th}
}

@article{Tekin2016,
  author = {Tekin, Bayram},
  title = {Particle Content of Quadratic and {$f(R_{\mu\nu\sigma\rho})$} Theories in {$(A)dS$}},
  journal = {Phys. Rev. D},
  volume = {93}, pages = {101502}, year = {2016},
  doi = {10.1103/PhysRevD.93.101502},
  eprint = {1604.00891}, archivePrefix = {arXiv}, primaryClass = {hep-th}
}

@article{LiLiuLu2018,
  author = {Li, Yue-Zhou and Liu, Hai-Shan and L{\"u}, H.},
  title = {Quasi-Topological {Ricci} Polynomial Gravities},
  journal = {J. High Energy Phys.},
  volume = {2018}, number = {2}, pages = {166}, year = {2018},
  doi = {10.1007/JHEP02(2018)166},
  eprint = {1708.07198}, archivePrefix = {arXiv}, primaryClass = {hep-th}
}

@article{Donoghue1994,
  author = {Donoghue, John F.},
  title = {General Relativity as an Effective Field Theory: The Leading Quantum Corrections},
  journal = {Phys. Rev. D},
  volume = {50}, pages = {3874--3888}, year = {1994},
  doi = {10.1103/PhysRevD.50.3874},
  eprint = {gr-qc/9405057}, archivePrefix = {arXiv}
}

@article{BeltranJimenez2019,
  author = {Beltr{\'a}n Jim{\'e}nez, Jose and Delhom, Adria},
  title = {Ghosts in Metric-Affine Higher Order Curvature Gravity},
  journal = {Eur. Phys. J. C},
  volume = {79}, number = {8}, pages = {656}, year = {2019},
  doi = {10.1140/epjc/s10052-019-7149-x},
  eprint = {1901.08988}, archivePrefix = {arXiv}, primaryClass = {gr-qc}
}

@article{DeserWaldron2001,
  author = {Deser, S. and Waldron, A.},
  title = {Partial Masslessness of Higher Spins in {$(A)dS$}},
  journal = {Nucl. Phys. B},
  volume = {607}, pages = {577--604}, year = {2001},
  doi = {10.1016/S0550-3213(01)00212-7},
  eprint = {hep-th/0103198}, archivePrefix = {arXiv}
}

@misc{Maldacena2011,
  author = {Maldacena, Juan},
  title = {{Einstein} Gravity from Conformal Gravity},
  year = {2011},
  eprint = {1105.5632}, archivePrefix = {arXiv}, primaryClass = {hep-th},
  doi = {10.48550/arXiv.1105.5632}
}

@article{Stelle1977,
 author={Stelle, K. S.}, title={Renormalization of Higher-Derivative Quantum Gravity},
 journal={Phys. Rev. D}, volume={16}, pages={953--969}, year={1977},
 doi={10.1103/PhysRevD.16.953}}

@article{Stelle1978,
 author={Stelle, K. S.}, title={Classical Gravity with Higher Derivatives},
 journal={Gen. Relativ. Gravit.}, volume={9}, pages={353--371}, year={1978},
 doi={10.1007/BF00760427}}

@article{Lovelock1971,
 author={Lovelock, David}, title={The Einstein Tensor and Its Generalizations},
 journal={J. Math. Phys.}, volume={12}, pages={498--501}, year={1971},
 doi={10.1063/1.1665613}}

@misc{Woodard2015,
 author={Woodard, R. P.}, title={The Theorem of Ostrogradsky}, year={2015},
 eprint={1506.02210}, archivePrefix={arXiv}, primaryClass={hep-th},
 url={https://arxiv.org/abs/1506.02210}}

@article{Biswas2012,
 author={Biswas, Tirthabir and Gerwick, Erik and Koivisto, Tomi and Mazumdar, Anupam},
 title={Towards Singularity- and Ghost-Free Theories of Gravity},
 journal={Phys. Rev. Lett.}, volume={108}, pages={031101}, year={2012},
 doi={10.1103/PhysRevLett.108.031101}, eprint={1110.5249}, archivePrefix={arXiv}}

@article{Starobinsky1980,
 author={Starobinsky, A. A.}, title={A New Type of Isotropic Cosmological Models Without Singularity},
 journal={Phys. Lett. B}, volume={91}, pages={99--102}, year={1980},
 doi={10.1016/0370-2693(80)90670-X}}

@article{MukhanovFeldmanBrandenberger1992,
 author={Mukhanov, V. F. and Feldman, H. A. and Brandenberger, R. H.},
 title={Theory of Cosmological Perturbations},
 journal={Phys. Rep.}, volume={215}, pages={203--333}, year={1992},
 doi={10.1016/0370-1573(92)90044-Z}}

@article{KofmanLindeStarobinsky1997,
 author={Kofman, Lev and Linde, Andrei and Starobinsky, Alexei A.},
 title={Towards the Theory of Reheating After Inflation},
 journal={Phys. Rev. D}, volume={56}, pages={3258--3295}, year={1997},
 doi={10.1103/PhysRevD.56.3258}}

@article{BanadosFerreira2010,
 author={Ba{\~n}ados, M{\'a}ximo and Ferreira, Pedro G.},
 title={Eddington's Theory of Gravity and Its Progeny},
 journal={Phys. Rev. Lett.}, volume={105}, pages={011101}, year={2010},
 doi={10.1103/PhysRevLett.105.011101}}

@article{BeltranJimenez2018,
 author={Beltr{\'a}n Jim{\'e}nez, Jose and Heisenberg, Lavinia and Olmo, Gonzalo J. and Rubiera-Garcia, Diego},
 title={Born--Infeld Inspired Modifications of Gravity},
 journal={Phys. Rep.}, volume={727}, pages={1--129}, year={2018},
 doi={10.1016/j.physrep.2017.11.001}, eprint={1704.03351}, archivePrefix={arXiv}}

@article{Higuchi1987,
 author={Higuchi, Atsushi}, title={Forbidden Mass Range for Spin-2 Field Theory in de Sitter Spacetime},
 journal={Nucl. Phys. B}, volume={282}, pages={397--436}, year={1987},
 doi={10.1016/0550-3213(87)90691-2}}

@article{BreitenlohnerFreedman1982,
 author={Breitenlohner, Peter and Freedman, Daniel Z.},
 title={Stability in Gauged Extended Supergravity},
 journal={Ann. Phys.}, volume={144}, pages={249--281}, year={1982},
 doi={10.1016/0003-4916(82)90116-6}}

@article{LuPangPope2011,
 author={L{\"u}, H. and Pang, Yi and Pope, C. N.},
 title={Conformal Gravity and Extensions of Critical Gravity},
 journal={Phys. Rev. D}, volume={84}, pages={064001}, year={2011},
 doi={10.1103/PhysRevD.84.064001}, eprint={1106.4657}, archivePrefix={arXiv}}

@article{DeserEtAl2011,
 author={Deser, S. and Liu, Haishan and L{\"u}, H. and Pope, C. N. and Sisman, Tahsin Cagri and Tekin, Bayram},
 title={Critical Points of {$D$}-Dimensional Extended Gravities},
 journal={Phys. Rev. D}, volume={83}, pages={061502}, year={2011},
 doi={10.1103/PhysRevD.83.061502}, eprint={1101.4009}, archivePrefix={arXiv}}

@article{GulluSismanTekin2011,
 author={Sisman, Tahsin Cagri and Gullu, Ibrahim and Tekin, Bayram},
 title={All Unitary Cubic Curvature Gravities in {$D$} Dimensions},
 journal={Class. Quantum Grav.}, volume={28}, pages={195004}, year={2011},
 doi={10.1088/0264-9381/28/19/195004}, eprint={1103.2307}, archivePrefix={arXiv}}

@article{Nariai1951,
 author={Nariai, Hidekazu},
 title={On a New Cosmological Solution of Einstein's Field Equations of Gravitation},
 journal={Sci. Rep. Tohoku Univ. Ser. I}, volume={35}, pages={62--70}, year={1951},
 url={https://cir.nii.ac.jp/crid/1571417125948039040}}

@article{HaoZhao2017,
  author = {Hao, Xin and Zhao, Liu},
  title = {Ricci polynomial gravity},
  journal = {Phys. Rev. D}, volume = {96}, pages = {124034}, year = {2017},
  doi = {10.1103/PhysRevD.96.124034}
}

@article{Planck2018VI,
  author = {{Planck Collaboration} and Aghanim, N. and others},
  title = {{Planck} 2018 results. {VI}. Cosmological parameters},
  journal = {Astron. Astrophys.}, volume = {641}, pages = {A6}, year = {2020},
  doi = {10.1051/0004-6361/201833910},
  eprint = {1807.06209}, archivePrefix = {arXiv}
}

@article{Lesgourgues2011CLASS,
  author = {Lesgourgues, Julien},
  title = {The Cosmic Linear Anisotropy Solving System ({CLASS}). {I}: Overview},
  journal = {arXiv e-prints}, year = {2011},
  eprint = {1104.2932}, archivePrefix = {arXiv}, primaryClass = {astro-ph.IM}
}

@article{BlasLesgourguesTram2011,
  author = {Blas, Diego and Lesgourgues, Julien and Tram, Thomas},
  title = {The Cosmic Linear Anisotropy Solving System ({CLASS}). {II}: Approximation schemes},
  journal = {JCAP}, volume = {2011}, number = {7}, pages = {034}, year = {2011},
  doi = {10.1088/1475-7516/2011/07/034},
  eprint = {1104.2933}, archivePrefix = {arXiv}
}

@article{Bertotti1959,
  author = {Bertotti, B.}, title = {Uniform Electromagnetic Field in the Theory of General Relativity},
  journal = {Phys. Rev.}, volume = {116}, pages = {1331--1333}, year = {1959},
  doi = {10.1103/PhysRev.116.1331}
}

@article{Robinson1959,
  author = {Robinson, I.}, title = {A Solution of the Maxwell--Einstein Equations},
  journal = {Bull. Acad. Polon. Sci. Ser. Sci. Math. Astron. Phys.},
  volume = {7}, pages = {351--352}, year = {1959}
}

@article{NojiriOdintsov2011,
  author = {Nojiri, Shin'ichi and Odintsov, Sergei D.},
  title = {Unified Cosmic History in Modified Gravity: from {F(R)} Theory to Lorentz Non-Invariant Models},
  journal = {Phys. Rep.}, volume = {505}, pages = {59--144}, year = {2011},
  doi = {10.1016/j.physrep.2011.04.001},
  eprint = {1011.0544}, archivePrefix = {arXiv}, primaryClass = {gr-qc}
}

@article{HehlMcCreaMielkeNeeman1995,
  author = {Hehl, Friedrich W. and McCrea, J. Dermott and Mielke, Eckehard W. and Ne'eman, Yuval},
  title = {Metric-Affine Gauge Theory of Gravity: Field Equations, {Noether} Identities, World Spinors, and Breaking of Dilation Invariance},
  journal = {Phys. Rep.}, volume = {258}, pages = {1--171}, year = {1995},
  doi = {10.1016/0370-1573(94)00111-F},
  eprint = {gr-qc/9402012}, archivePrefix = {arXiv}
}

@article{BarausseSotiriouMiller2008,
  author = {Barausse, Enrico and Sotiriou, Thomas P. and Miller, John C.},
  title = {Curvature Singularities, Tidal Forces and the Viability of {Palatini} $f(R)$ Gravity},
  journal = {Class. Quantum Grav.}, volume = {25}, number = {10}, pages = {105008}, year = {2008},
  doi = {10.1088/0264-9381/25/10/105008},
  eprint = {0712.1141}, archivePrefix = {arXiv}, primaryClass = {gr-qc}
}

@misc{Baumann2009,
  author = {Baumann, Daniel},
  title = {{TASI} Lectures on Inflation},
  year = {2009},
  eprint = {0907.5424}, archivePrefix = {arXiv}, primaryClass = {hep-th},
  doi = {10.48550/arXiv.0907.5424}
}
\end{document}